\documentclass{aa}

\usepackage{graphics}

\begin{document}
\thesaurus{02 (12.03.1; 12.03.4)}  
\title{Distortion of the CMB Spectrum by Primeval Hydrogen Recombination}
\author{\sc P\'eter Bosch\'an and Peter Biltzinger}     


\institute{Institute for Nuclear Physics, University of M\"unster, 
Wilhelm--Klemm--Str. 9, 48149 M\"unster, Germany}

\date{Received 26 August 1997 / Accepted 27 April 1998}

\maketitle
\markboth{
{\sc P. Bosch\'an \& P. Biltzinger}: Distortion of the CMB 
Spectrum by Primeval Hydrogen Recombination}    
{{\sc P. Bosch\'an \& P. Biltzinger}: Distortion of the CMB 
Spectrum by Primeval Hydrogen Recombination}

\begin{abstract}
We solve the recombination equation by taking into account the induced 
recombinations and a physical cut off in the hydrogen spectrum. The effective 
recombination coefficient is parametrized as a function of  
temperature and free electron density and is about a factor of four larger than
without the induced recombination. This accelerates the last
stage of the recombination processes and diminishes the residual ionization
 by a factor of about 2.6. The number and energy distribution of photons
 issuing from the hydrogen recombination are calculated. The distortion of
 the cosmic microwave background (CMB) spectrum  depends strongly on
 the cosmological parameters
 $\Omega, \,\Omega_b \, {\rm and} \, H_0 \,$ and differs essentially
 from the Planck-spectrum for wavelengths $\leq 190 \mu m$. 
\end{abstract}


\section{Introduction}

The CMB radiation in a wide frequency range has a Planck--spectrum
with $T_0 = 2.728\pm 0.002 K$ (Fixsen et al.\ 1996). The last dramatic event
that could have influenced a part of the spectrum was the recombination 
of the primeval hydrogen, since after it the  residual ionization was
very low and  the radiation fields practically don't interact 
with the non-relativistic matter any more.  The properties of the 
post--recombination Universe can be studied by observing the microwave
background radiation.
The fluctuations in the CMB are supposed to be the precursors of the 
largest structures observed today. 

To understand the nature of the microwave background anisotropies it is 
necessary to have a correct picture of the recombination process itself.
The problem was first studied shortly after the discovery of the CMB by
Peebles (1968) and at the same time by Zel'dovich, Kurt and Sunyaev (1968).
Since than several authors have adressed the problem, but no one could 
improve on the basic approximations used in these works. However many details were worked out.
Matsuda, Sato \& Takeda (1971) examined the effects of the
collision processes and found that they are negligible with respect to the
radiation processes.
Jones \& Wyse (1985) improved on the calculations to allow for the presence of 
non--baryonic matter. Krolik (1989, 1990) showed that two previously equally 
neglected scattering effects in the Ly$_\alpha$ line almost completely 
cancel each other. Sasaki \& Takahara (1993) concluded that taking
induced recombination into account changes the time history of the 
recombination at very low ionization grades, and lowers the residual
ionization. They however used an approximation for the recombination cross 
sections and did not discuss the divergence of the sum of the recombination 
rates.
Dell'Antonio \& Rybicki (1993) followed the progress of recombination 
numerically, using a multilevel hydrogen atom  model; all angular momentum 
states were treated individually up to $n = 10$. Their recombination curve was
similiar to that of  Jones \& Wyse (1985).
In an other article
Rybicki \& Dell'Antonio (1994) calculated the time-dependence of the 
Ly$_{\alpha}$ line profile in a homogenious expanding Universe. They found 
that the usual quasi-static approximation for the line shape
is reasonable and does not cause substantial error in the solutions. 
They also determined the recombination history for several cosmologies.
Hu, Scott, Sugiyama \& White (1995) discussed the
effect of physical assumptions for CMB anisotropies and on the recombination
process.

Recombination, as it has been pointed out in the papers quoted above, sets in 
when all energy levels of hydrogen, except for the ground state are still in 
equilibrium with radiation. When one hydrogen atom enters the  ground 
state, a photon with energy greater than or equal to the energy 
of a ${\rm Ly_{\alpha}}$ transition
is emitted. Photons, capable of exciting electrons in the ground state, 
are lost via two 
main routes: 1. by cosmological redshift of ${\rm Ly_{\alpha}}$ photons, 
2. by two--photon transition $2s \rightarrow 1s$. Of these two competing 
processes the first is purely cosmological, the second is atomic. 
Their relative importance depends on the cosmology.

By theoretical examination of the recombination process one computes the
time dependence of the ionization rate, the residual ionization, the position
and width of the last scattering layer and the distortion of the CMB spectrum. 
Though the amount of the residual ionization is important for the further
evolution of the Universe (Peebles 1993, Lepp \& Shull 1984), it can  
not be measured directly. A possible
observable consequence of the hydrogen recombination in the early Universe 
is the distortion of the microwave background radiation spectrum. 
It was first calculated for the flat cosmological model by Peebles (1968). In 
this work we concentrate on the determination of this distortion for different
cosmological models. 
Recently there has been a considerable observational and theoretical activity 
to determine the spectrum of the cosmic background radiation and carrying out
galaxy counts in the far infrared/submillimeter range (Puget et al. 1996, 
Schlegel et al. 1997, Guideroni et al. 1997). 
Because the hydrogen recombination changes the CBM spectrum in the same 
spectral range a detailed exact recalculation of the frequency distribution 
of the recombination photons is important. In this paper we calculate
 the recombination cross sections exactly by using a continuous, physical cut
 off for the highly excited states of hydrogen, take into account
 the induced recombination and explain, why the time dependence of the
 recombination process is so hardly effected by different techniques and
 by different effective ionization curves.

The outline of the paper is as follows. In \S 2 we discuss the recombination
process and derive the recombination equation.  We give a new 
parametrization of the effective recombination coefficients, taking into 
account the induced recombination as well. In the third section 
we solve the recombination equation. The spectrum of the recombination photons
and its dependence on the cosmological parameters is given in the fourth \S ,
and in the fifth closing section we discuss our results.

\section{The Recombination Process}

\subsection{The Excited States of Hydrogen}

The  number of recombinations in unit time can be calculated from the
recombination equation as 
a function of the density  of free electrons $n_e\,$, temperature $T\,$, and 
cosmological, and atomic constants. In this section we derive 
this equation  following Peebles (1968).

At the beginning of the hydrogen recombination the helium is already 
completely recombined. The mass fraction of the helium is 25 \% of the 
total baryonic mass.
As was remarked by Novikov and Zel'dovich (1967) the direct recombinations
to the ground state are inhibited, while the new born energetic photons ionize
again
almost immediately when there are already some hydrogen atoms.  In the 
following we neglect completely the direct recombinations to the ground 
state.

The states with principal quantum number $n = 2 \,$ play a key role  in
the recombination processes. First we calculate the rate of 
recombinations to $ n \geq 2\,$ for given free electron number density 
$n_e$, temperature T and $n_{2s}$.
Here $n_{nl}$ is the density of atoms in the state with the principal quantum 
number $n$ and angular momentum quantum number $l$. Second, we determine 
the net number of $2 \rightarrow 1$ transitions in unit time by given 
$n_{2s} $ and $n_{1s}$.
The two rates must be equal, so we can eliminate $n_{2s}$ from the 
calculations.

The binding energy of the $n = 2 \,$ state is B$_2 = {\rm B}_1/4 \approx 
3.4\, eV$.
During the recombination process there are a large
number of photons with energy less than $B_{2}$, therefore the excited states 
of the
hydrogen atoms are in thermodynamical equilibrium above the second  level, i.e.
\begin{eqnarray}
        n_{nl}^{({\rm Boltzmann})} &=& {(2 l +1) \over 4} e^{-{B_2 - B_n 
        \over k T }} n_2 \nonumber \\ 
         &=&  (2 l +1)  e^{-{B_2 - B_n \over k T }} n_{2s}.
\end{eqnarray}
In the case of gaseous nebulae the mean free path of low energetic photons
is larger than the dimensions of the ionized region, the system is far from
being in  equilibrium  above the $ n = 2\,$ level in contrast to the 
recombining Universe.

The partition sum $\sum_{nl} (2 l + 1) n_{nl}^{({\rm Boltzmann})} \,$ 
is divergent. As analysed
by Hummer and Mihalas (1988) a number of effects limit the range of summation.
In the considered temperature and density range the action of free protons
turns out to be the most important, the action of neutral atoms are much 
smaller. The free protons destroy the state $n\,$ with a 
probability $ 1 - w_n\,$, where $w_n = \exp[\,-({n \over n_*})^{15/2}\,]\,$,
with $n_* = 1075.9 \,{\rm cm}^{-2/5}\,  n^{-2/15}_e\,$ (n$_e$ in cm$^{-3}$). 
This means that the recombined electrons become unbound
with $1- w_n$ probability, before they would begin to move towards the state,
corresponding to thermal equilibrium. 
The highly excited ($n$ larger than $\sim n_*$) states are practically
completely destroyed. The occupation numbers of the excited states in thermal
equilibrium are:
\begin{equation} 
\label{nnl}
        n_{nl} = 
         (2 l +1)  e^{-{B_2 - B_n \over k T }} n_{2s} \, w_n. 
\end{equation} 
Owing to this fact, the number of electrons in bound states is finite, and
thermal equilibrium between the $n \geq 2$ bound states and the continuum
is possible.
This approximation is good down to z $\approx \,$ 300. The $n = 2 \,$ levels
freeze out between the redshifts 300 and 250.
  
The ground state of the hydrogen atom is about 10.2 eV deeper than the $n = 2$ 
level. Consequently, it's occupation is greater than the occupation of the 
excited states together: $ n_{1s} \gg \sum_{n \geq 2} n_n$. This means, that 
the
density of free electrons plus the density of hydrogen atoms in ground state
can be taken as equal to the total proton number : $ n_e + n_{1s} = p\,$.
The total baryon number (determined by $\Omega_b$) is the sum of p and
the number of baryons in the helium nuclei.

\subsection{Recombination on Excited States of Hydrogen}

The number of recombinations to the level $nl$ and the number of ionizations
from $nl$ in unit time are
\begin{equation}
\label{rec}
      \alpha_{nl}\, n_e\, n_p = \alpha_{nl}\, n_e^2 \quad\quad {\rm and} \quad 
        \quad\beta_{nl}\, n_{nl}                    
\end{equation}
where $\alpha_{nl}$ and $\beta_{nl}$ are the recombination and ionization 
coefficients, $n_p$ is the density of free protons. The usual definition of
these quantities are 
\begin{equation}
\label{ion}
        \alpha_{nl}^{\prime} = < \sigma_{nl}^{({\rm rec})} v>_{{\rm Maxwell}},
         \quad  \beta_{nl}  = 
        < \sigma_{nl}^{({\rm ion})} c>_{{\rm Planck}}.  
\end{equation}
Here $v$ and $c$ are the velocities. The averages must be taken over 
the Maxwell--distribution for electrons and the Planck--distribution for 
photons.

The ionization coefficients are:
\begin{equation}
\label{cion}
        \beta_{nl} = (2 l + 1) \int c\, \sigma^{(ion)}_{nl} (\nu) 
        {8 \pi \nu^2 \over c^3}{2 \over e^{ {h \nu \over k T}} -1 } d \nu \,.
\end{equation}
The relation between the recombination and ionization cross sections is: 
$ \sigma_{nl}^{(rec)} = \sigma_{nl}^{(ion)} {2 {\bf k}^2 \over {\bf p}^2}\,$,
where ${\bf k\,}$ and  ${\bf p}\,$ are the photon and electron momenta.

By using, instead of the electron velocity, the energy of the recombination 
photon in the Maxwell--distribution and expressing the recombination cross 
section through the ionization cross section, one arrives at
\begin{eqnarray}
\label{rcoefo}
\alpha_{nl}^{\prime}(T)&=&{4 \pi  2 (2 l+1)\over c^2 
(2 \pi m_e k T)^{{3 \over 2}}} e^{{B_n \over k T}}\!  \int_{B_n}^{\infty}
\sigma_{nl}^{(ion)} (h\nu)^2  e^{-{h \nu  \over k T}}  d \, h \nu ,
\end{eqnarray}
where $m_e$ is the electron mass.
The integrals in $\alpha_{nl}^{\prime}$ and $\beta_{nl} \,$ are not the same.
But, due to the principle of detailed balance (Milne 1924)  the recombinations 
and ionizations  in thermodynamic equilibrium must exactly cancel each other.
If we let contributions from the induced recombinations modify the 
recombination coefficients (e.g. Mihalas 1984, Sasaki \&  Takahara 1993)
instead of (\ref{rcoefo}) we get
\begin{eqnarray}
\label{rcoef}
\alpha_{nl}(T)&=&{4 \pi \, 2 (2 l+1)\over c^2 (2 \pi m_e k T)^{{3 \over 2}}}
e^{{B_n \over k T}} \int_{B_n}^{\infty}
\sigma_{nl}^{(ion)}{ (h\nu)^2 \over e^{{h \nu  \over k T}} - 1} d \, h \nu \nonumber \\
&=&  {\Bigl({2 \pi \hbar^2 \over  m_e k T} \Bigr)}^{3/2} e^{{B_n \over k T}} 
\beta_{nl} .             
\end{eqnarray}
The term $ -1 \,$ under the integral in the 
denominator comes from the induced recombinations. Without it the principle of
detailed balance would not be fulfilled. 

Summing up over all $n,l$ levels and making use of equation (\ref{nnl}) the 
total rate of recombinations is:
\begin{equation}
\label{recr}
        \alpha n^2_e \quad\quad {\rm with} \quad \alpha = \sum^{\infty}_{n=2}
        \sum_{l=0}^{n-1} \alpha_{nl} \, w_n.        
\end{equation}
The total rate of ionizations from the excited states of hydrogen is:
\begin{equation}
\label{ionr}
        \sum^{\infty}_{n=2} \sum_{l=0}^{n-1} \beta_{nl}\, n_{nl} \,w_n
        = \Bigl({m_e k T \over 2 \pi \hbar^2} \Bigr)^{3/2} e^{-{B_2 \over k T}}
        \alpha \, n_{2s} = \beta n_{2s} .                    
\end{equation}
The net number of recombinations to excited levels in unit time and unit 
volume is the difference of these two expressions. The number density of the
electrons changes, not only because of the recombination process, but also in 
consequence to the universal expansion. The simplest way to take  this effect
into account is to write on
equation for the ionization rate, i.e.\ the ratio 
of free electrons to  the number of free protons plus hydrogen atoms  :
\begin{equation} 
\label{rec2s}
        -{d  \over d t} {n_e \over p} = \alpha\, \Bigl[{n_e^2 \over p} -
        \Bigl({m_e k T \over 2 \pi \hbar^2} \Bigr)^{3/2} e^{-{B_2 \over k T}}
        {n_{2s} \over p}\Bigl].                     
\end{equation}

\subsection{Depopulation of the $n = 2 $ level}

Each $2p \rightarrow 1s$ transition liberates a ${\rm Ly}_{\alpha}$ photon,
which can excite the ground state, and there are always enough low energy 
photons to ionize from an excited state.
The transition to the ground state is final, when the
number of these $E_{\gamma} \ge B_1 - B_2 \,$ photons is also diminished. 
Photons can escape from the   ${\rm Ly}_{\alpha}$ line either by redshift or
by two--photon  2s $\rightarrow$ 1s  transitions.


In a stationary state the numbers of emitted and absorbed 
${\rm Ly}_{\alpha}\,$ photons are equal. 
Consequently, the number of photons per mode in the
${\rm Ly}_{\alpha}$ resonance line is:
\begin{equation}
\label{gamm}
        \gamma_{12} = {n_{2p} \over 3 n_{1s} - n_{2p}} \approx {n_{2p} 
        \over 3 n_{1s}} = {n_{2s} \over n_{1s}}. 
\end{equation}
Because of the general expansion of the Universe a frequency $\nu_0$ is 
shifted in unit time by
\begin{equation}
\label{reds}
        {\delta \nu_0 \over \delta t} = \nu_0 H ,  
\end{equation}
where H is the Hubble--parameter. The number of recombination photons
removed from the line by redshift in unit time is:
\begin{eqnarray}
\label{rsr}
        R_{{\rm rs}} &=&{\delta \nu_{12} \over \delta t}
        {8 \pi  \nu_{12}^2 \over c^3}
        {\bigl( \gamma_{12} - \gamma_{12} 
        ({\rm background}) \bigr)}\nonumber \\
        &=& H {8 \pi \nu_{12}^3 \over c^3}
         {\bigl( \gamma_{12} 
        - {1 \over e^{h \nu_{12} /k T} -1} \bigr)}.
\end{eqnarray}

On the other hand, the rate of two--photon decay in the 2s states has been
calculated by Spitzer \& Greenstein (1951). The number of net decays in unit
time is
\begin{equation}
\label{rph2}
        R_{2 {\rm phot}} = A_{2s,1s} \bigl(n_{2s} - n_{1s} e^{-(B_1 - B_2 )
        /k T} \bigr),                                   
\end{equation}
with $A_{2s,1s} = 8.227 s^{-1}$. In thermal equilibrium the numbers of 
two--photon decays and the two--photon $1s \rightarrow 2s\,$ excitations are 
equal.
 
At this point we have four unknown quantities: $ n_{1s},n_{2s}, \gamma_{12}
\,{\rm and}\, n_e \,$ and four equations (\ref{rec2s}), (\ref{gamm}), 
(\ref{rsr}) and (\ref{rph2}); so we are left with
\begin{eqnarray}
\label{receq}
        -{d \over dt} \,{n_e \over p} &&= \Bigl[ { \alpha n_e^2 \over p} - 
        \beta \bigl( 1 - {n_e \over p} \bigr) e^{-(B_1 - B_2 )/k T } 
        \Bigr] \nonumber \\
&& \cdot {8 \pi \nu^3_{12}  c^{-3} H + A_{2s,1s} (p - n_e )
        \over 8 \pi \nu^3_{12} c^{-3} H + (A_{2s,1s} + \beta)  (p - n_e )}.
\end{eqnarray}

The time dependence of $p,\, T \, {\rm and} \,H $ are calculated from the 
Friedmann equation with the density parameter $\Omega  (\, = \Omega_b + 
\Omega_{\gamma} + \Omega_{DM})$, the baryon density parameter 
$\Omega_b \,$ and the Hubble constant $ H_0 = 100 \, {\rm h \,km \,s}^{-1} 
{\rm Mpc}^{-1}$ ($\Lambda = 0$, the cosmological constant).

\subsection{Parametrization of the recombination coefficient}

To solve the equation (\ref{receq}) we need the recombination coefficient
$\alpha $, wich is parametrized as a function of the temperature in several
ways. Peebles (1968) used $\alpha = 2.84\cdot 10^{-11}/ \,\sqrt{T} 
{\rm cm^3\, s^{-1}} \,$ on the basis of Boardman's (1964) data for the seven 
lowest states. Zel'dovich, Kurt and Sunyaev (1968) used 
$ 2.5\cdot 10^{-11}/ \sqrt{T}\,{\rm cm^3\, s}^{-1}\,$.
Peebles (1993) using Osterbrock's (1989) data got the
parametrization:
$\alpha = 4.1\cdot 10^{-10}\, T^{-0.8}\, {\rm cm^3\, s^{-1}}$. 
Recently  Hummer (1994) tabulated very accurately the recombination rates.
His results confirm  the parametrization given by P\'equignot et al. (1991)
in the form of 
\begin{equation}
\label{alpeq}
        \alpha(T) = 1.261 \cdot 10^{-10} {T^{-0.6166} \over 1 +
        5.087  \cdot 10^{-3} T^{0.53}}.
\end{equation}
Sasaki \& Takahara (1993) had used  the asymptotic form for the recombination 
cross sections which, also  valid for large principal quantum numbers $n$,
was less than 20\% accurate for the important low lying levels.

When calculating the recombination coefficients without taking into account
the induced recombination, the $\,- 1\,$  falls out of  the denominator 
of the integral in (\ref{rcoef}). 
In this case the $\alpha_n \,$ series converges as
$ {\rm log}(n)/n^3 $, so one can neglect the contribution of the terms with 
large $n$. 
If induced recombination is taken into account,  the denominator under the 
integral in (\ref{rcoef}) goes to zero  for large $n$, and the series
of the $\alpha_{n}$ diverges as $1/n \,$.

Here we calculate the total recombination coefficient from (\ref{recr}), 
 wherein the factor $ { w_n}$, the probability that the states 
with principal
quantum number $n$ are not destroyed, makes the sum finite. The probability
that a state is destroyed depends on the baryon density of the Universe. 
Consequently, the effective recombination coefficient also density dependent.

Our procedure is then the following. We calculate the first several 
hundred $\alpha_{nl}$'s using the  exact results given by
Storey \& Hummer  (1991). For large $n$ ($ \ge 200 \,$), when the recombination
coefficients have already reach their asymptotic values (the difference 
between 
the exact and asymptotic value $ < 0.1\% \,$), we use the 
 $\alpha_n $'s given e.g.\,\,in Rybicki \& Lightman (1979). The density dependence comes in through
the $w_n \,$ probabilities. The effective recombination  coefficient is 
parametrized  as a function of the temperature, $T$, and the electron number
density, ${ n_e}$, in the form
\begin{eqnarray}
\label{alpar} \nonumber
\alpha({T,n_e}) &=& 1.21\cdot 10^{-10}\, T^{-{1 \over 2}} \\
 && \cdot \,(\,1 - 0.0226 \,\,{{\rm ln} (n_e\, cm^3) }\,\,)\,\, {\rm cm}^3 \,\,
 {\rm s}^{-1}\, .           
\end{eqnarray}
For a given temperature $n_e$ and $\alpha$  depend on the cosmological 
parameters. However this dependence is mild, not larger than
10 \% for the usual range of parameters.
We compute the $\alpha ({T,n_e})\,$ function together with $n_e$ by solving 
(\ref{receq}). The solution of this equation for n$_e$ and 
$ \alpha({ T,n_e}) \,$ will be given in the next 
section.  For the sake of clarity we present our recombination rate
already here  (Fig. 1). The form
 of the $ \alpha({ T,n_e(T)}) \,$ function is not far from the usual 
$\sim T^{-0.7}$ function, but its  amplitude is about four times larger than 
it would be without taking the induced recombination  into account. 
Furthermore we compare our $\alpha (T) \,$  with  (\ref{alpeq}), and
with Sasaki \& Takahara (1993), whereby our function is somewhat steeper 
than that of these authors.

\begin{figure}[htbp]
\resizebox{\hsize}{6cm}{\includegraphics{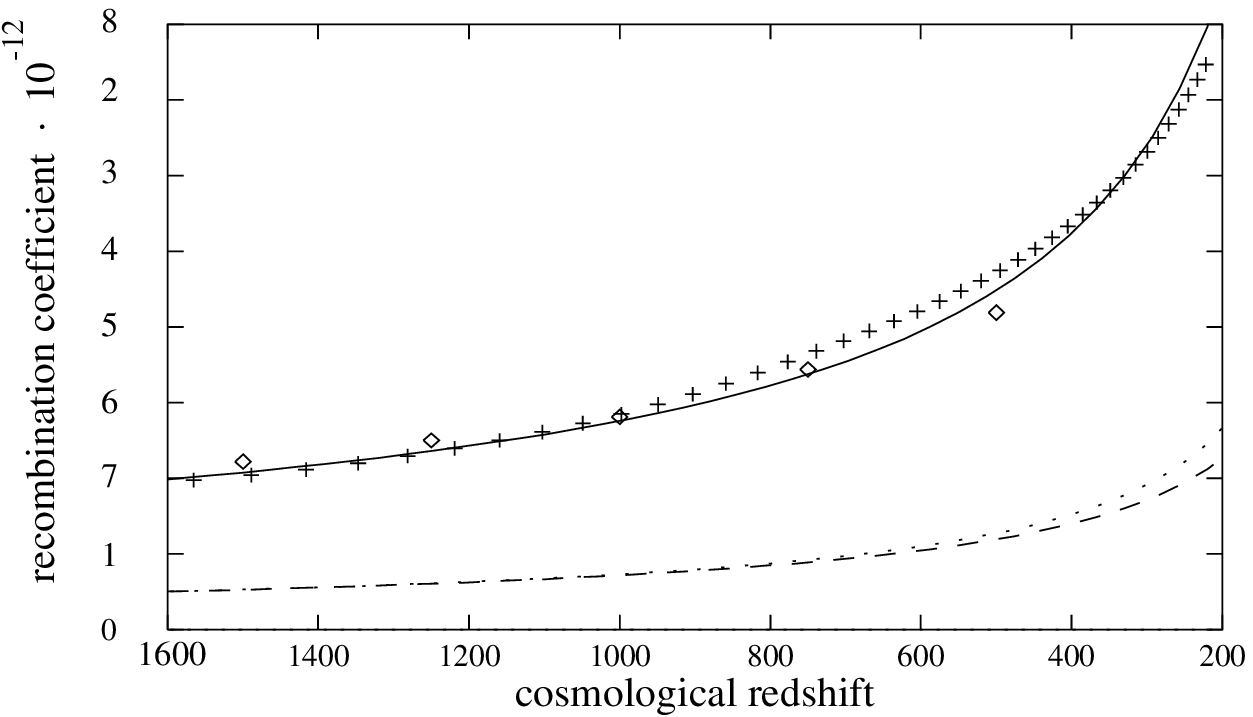}}
\caption{Recombination coefficient as a function of redshift.}
Crosses: Present work (for $\Omega = 1,\, \Omega_b=0.1, \,$ h = 1);
Continuous line:  $T^{-0.7}$ fit; Diamonds:  Sasaki \& Takahara 1993;
Points: eq. (\ref{alpeq}); Dashed line:  Peebles 1993.
\label{Fig1}
\end{figure}


\section{The solution of the recombination equation}

We solved the equation (\ref{receq}) for recombination numerically using 
equations (\ref{alpar}) and (\ref{ionr}). The results with and 
without taking into account the  induced recombination are  compared.

\subsection{The residual ionization}
\begin{table}
      \caption{Residual ionization at $z=220$ for five cosmolgical models.
$x_2 \,$ and $x_1 $ are the residual ionizations calculated 
with and without taking into account the induced recombination.} 
\label{table1}
\begin{flushbottom}       
      \begin{tabular}{llll}
            \hline \noalign{\smallskip}
            $\Omega$ & 1   & 1 & 0.5 \\
            \noalign{\smallskip}
             $\Omega_b$ & 1   & 0.06   & 0.1 \\
            \noalign{\smallskip}
            $h$ & 1   & 1   & 1  \\
            \hline \noalign{\smallskip}
             $x_1$&$2.1359 \cdot 10^{-5}$ & $3.4811 \cdot 10^{-4}$ &
             $1.5068\cdot10^{-4}$\\
             $x_2$ & $8.3545 \cdot 10^{-6}$& $1.3565\cdot 10^{-4}$
             &$5.8440\cdot10^{-5}$\\
             $x_1/x_2$ & 2.5567 & 2.5662& 2.5784\\
            \noalign{\smallskip}  \hline
         \end{tabular}\\ \\ \\
      \begin{tabular}{lll}
            \hline \noalign{\smallskip}
            $\Omega$ &  0.1  & 0.3\\
            \noalign{\smallskip}
             $\Omega_b$ & 0.1  & 0.01\\
            \noalign{\smallskip}
            $h$ & 0.75  & 0.5 \\
            \hline \noalign{\smallskip}
             $x_1$&$ 9.9243 \cdot 10^{-5} $ & $2.2974 \cdot 10^{-3}$\\
             $x_2$&$ 3.7740\cdot 10^{-5}$ & $8.6741\cdot 10^{-4}$\\
             $x_1/x_2$ & 2.6297& 2.6486 \\
            \noalign{\smallskip}  \hline
         \end{tabular}
        \end{flushbottom}	
\end{table}


 At $z \sim$ 200 ${\rm H_2}\,$ formation becomes possible. Since
free electrons and protons serve as catalysts for the formation of molecular
hydrogen (Peebles 1993), the value of the ionized fraction at this epoch 
is a very important. 

At about $z \sim$ 500  $\beta$ becomes much smaller than A$_{\rm 2s,1s}$. It is
independent of the cosmology, because both $\beta$ and A$_{2s,1s}$ are atomic
 quantities. In this 
case the fraction in equation (\ref{receq}) is one. 
The second term in the square bracket is also small, so instead of the equation
(\ref{receq}) for $x = n_e /{\rm p}$ we can write 
\begin{equation}
\label{assol}
        {d x \over dz } = {\alpha p \over z H} x^2 = {\rm const} 
        {\sqrt{\Omega} \over \Omega_b h} x^2 (1 - 0.0226\, {\rm ln}\,  p\, x).
\end{equation}
When the induced recombinations are not taken into account the density
dependent term does not appear and the residual ionization scales with 
$\sqrt \Omega \Omega_b^{-1} h^{-1}\,$. 
The logarithmic term contains another combination of the cosmological 
parameters, therefore the scaling law is mildly violated.
The residual ionizations at $z = 220\, $ are shown in Table 1. 
When the induced recombination is taken into account the residual ionization
is reduced by  a factor of about 2.6\,. Small deviations from this factor
are due to the violation of the scaling law. The total number of
 ions is proportional to $x \,  \Omega_b$, so it  is 
almost independent of the baryon density.

We refined the previous treatment (Sasaki\& Takahara 1993) of the induced 
recombination by using the exact recombination cross sections and applying
a physical cutoff in the summation on the hydrogen states. This causes a
10 - 20 \% change in the residual ionization.

\subsection{The time dependence of the fractional ionization } 

The recombination history for different sets of comological parameters is
  shown in Fig.\ 2. The quotient of the fractional ionization at different cosmologies grows with
time and approximately follows the x $\sim \sqrt{\Omega} \Omega_b^{-1} 
h^{-1}\,$ scaling law. We compare our result with Dell'Antonio and
Rybicki (1993), who did not take  induced recombinations into account. 
For corresponding parameters our recombination curve runs significantly 
under their curve, because the last phase of the recombination processes is 
accelerated by the induced recombinations.
\begin{figure}[htbp]
\resizebox{\hsize}{8cm}{\includegraphics{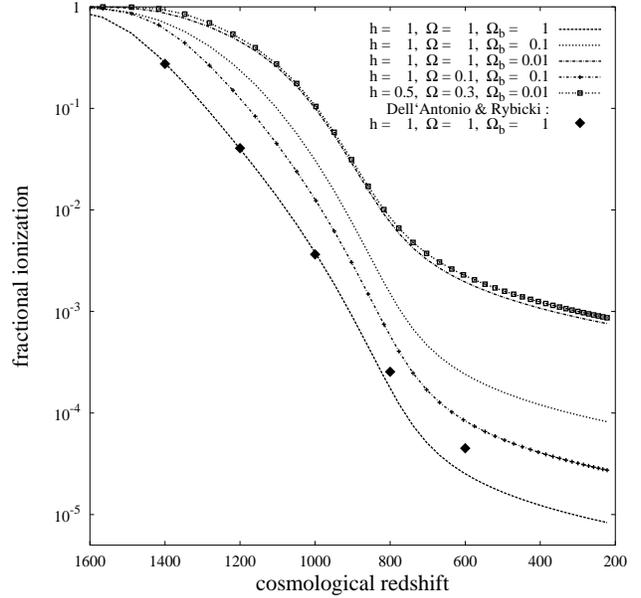}}
\caption{Recombination history for different cosmological models.}
\label{Fig2}
\end{figure}

\begin{figure}[htbp]
\resizebox{\hsize}{8cm}{\includegraphics{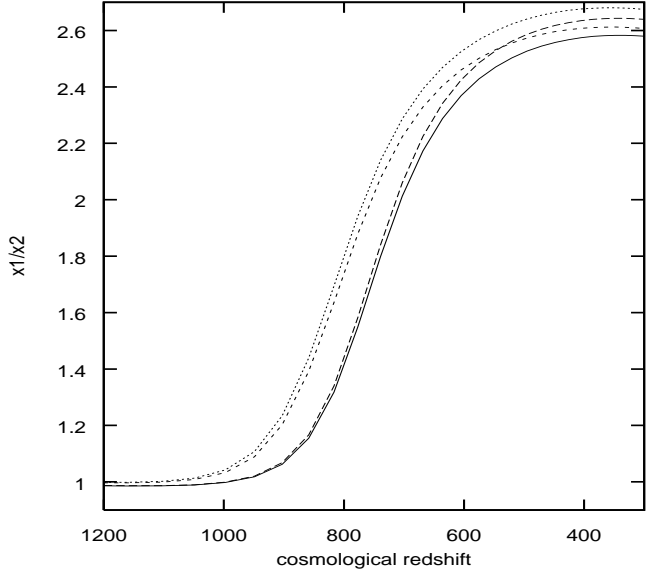}}
\caption{Ratio of the fractional ionization calculated 
with ($x_2$) and without ($x_1$) taking  induced recombination into account. 
From right to left:}
\vspace*{-0.3cm}
\begin{tabbing}
a)\qquad \= $\Omega $ = 1,\= \qquad $\Omega_b$ = 1,\= \qquad  h = 1  \\
b) \qquad \> $\Omega $ = 0.1,\> \qquad  $\Omega_b$ = 0.1,\> \qquad h = 1 \\
c) \qquad \> $\Omega $ = 1, \> \qquad  $\Omega_b$ = 0.01,\> \qquad h = 1 \\
d)\qquad  \>$\Omega $ = 0.3,\>  \qquad $\Omega_b$ = 0.01,\> \qquad h = 0.5
\end{tabbing}
\label{Fig3}
\end{figure}

In Fig.\ 3 we show the ratio of fractional ionization calculated with 
($x_2\,$) and without ($x_1) \,$ taking  induced recombination into account.
In the first part of the process the induced free--bound 
transitions and the numerical value of the effective 
recombination coefficient have little effect on  the course of the process,
because the depopulation of the $n = 2$ states is much slower than the 
recombination to the excited
 states, the thermal equilibrium corresponding to the 
occupation of $n = 2$ is always maintained. Down to $z \approx 1000$ the 
$x_1 / x_2 \,$ ratio is one. Between 
$z \approx 1000$ and $z \approx 500$ the change owing to induced 
recombinations depends on the cosmology. Later, when   $z \le 500$, 
especially when the baryon density $\Omega_b \,$ is small,
the depopulation of $n = 2$ states is faster than the recombination to the
excited states and this last process determines the net recombination rate. 
At this stage the induced recombinations play an important role. As can
be seen in Fig.\ 3, the $x_1/x_2\,$ ratio depends hardly on the cosmological 
parameters.

\subsection{The last scattering surface}

From the ${ n_e (z)}\,$ function one can estimate the probability that
a CMB photon has not been scattered since a given redshift $z$. Instead of this
probability, usually its differential is calculated. This quantity determines
the probability density that the radiation was last scattered at $z$ and 
is expressed as $g(z) = e^{-\tau} \, d \tau / d z\,$. Here  $\tau \,$ is the 
Thomson scattering optical depth:
$ \tau = - \int_0^z  c \sigma_T n_e(z) {d t \over dz} d z \,$, and 
$\sigma_T \,$ is the Thomson cross section.
The parameters, determining the position and width of this layer are 
given in Table 2. These results are in reasonable agreement with those of
White, Scott \& Silk (1994). This occurs because the last Thomson scattering 
happens early on, when the induced recombinations don't play any role.

   \begin{table}
      \caption{Position and width of the last scattering surface, calculated
for different sets of cosmological parameters.}
\label{table2}
        \begin{flushleft}       
      \begin{tabular}{llllll}
            \hline \noalign{\smallskip}
            $\Omega$ & 1  & 1 & 1   & 0.1 & 0.1\\
            \noalign{\smallskip}
            $\Omega_b$ & 1   & 0.1  & 0.01 & 0.1  & 0.1\\
            \noalign{\smallskip}
            $h$ & 1 & 1  & 1   & 1 & 0.5\\
            \hline \noalign{\smallskip}
            $z_{\rm rec}$& $1032.0$ & $1043.6$&$1120.0$&$ 1028.4$
            &$1020.1$\\
            $\Delta z$&$85.7$& $ 90.8$&$134.3$&$85.0$ & $83.8$\\
            \noalign{\smallskip}
            \hline
         \end{tabular}
        \end{flushleft}
   \end{table}

\section{The spectrum of the recombination photons}

\subsection{The time variation of the photon spectrum}

The evolution of the photon spectrum during recombination can be calculated 
from the continuity equation in frequency space (Peebles 1968):
\begin{eqnarray}
\label{photeq}
{\partial \over\partial t} \left( {\nu\, n(\nu,t) \over p(t)} \right) &=&
H(t)\,\nu\,{\partial \over \partial \nu} \left( {\nu\,n(\nu,t) 
\over p(t)} \right) \nonumber \\ &+& {\nu\,\,Q(\nu,t) \over p(t)} \, .  
\end{eqnarray}
${ n(\nu ,t)}\,$ is the number density of photons with frequency 
$\nu $, and ${Q(\nu,t)} \,$ is the net rate of production of photons per 
unit volume and unit frequency interval. The Planck--spectrum fulfills this 
equation
without the source term, so (\ref{photeq}) is also true for 
${n^{\prime} (\nu,t) = n(\nu,t) - n_{{\rm Planck}}(\nu,t)}\,$.
Here we introduce new dependent variables with the definition
\begin{equation}
\label{geq}
        g(\nu,t) ={\nu  \over p(t)}\, n^{\prime}(\nu,t) 
\end{equation}
and define the  independent variables $x, \tau \,$ as
\begin{equation}
\label{vareq}
        x = \ln {h \nu \over B_1} \quad {\rm and} \quad 
        \tau = \int^{t} H(t^{\prime}) dt^{\prime}.
\end{equation}
Instead of the equation (\ref{photeq}) we have
\begin{equation}
\label{lapleq}
\left( {\partial \over \partial \tau} - {\partial \over \partial x} \right) 
g(x,\tau) = {B_1 \,e^x \over 2 \pi \hbar \,H(\tau)\, p(\tau)}\, Q(x,\tau).
\end{equation}
In the absence of sources the spectrum moves to the left 
(with ``velocity'' 1) in the $(\tau \, ,x )\,$ plane.

When the right hand side of (\ref{lapleq}) differs from zero, the solution  can be written
in integral form as:
\begin{equation}
\label{sollap}
        g(x,\tau) = \int^{x+\tau}_x {B_1 \,e^{x^{\prime}} \over 2 \pi \hbar,
        H(\tau^{\prime})\, p(\tau^{\prime})}\, Q(x^{\prime},\tau^{\prime}) 
        d x^{\prime}, 
\end{equation}
with $\tau^{\prime} = x + \tau - x^{\prime}$.

\begin{figure}[htbp]
\resizebox{\hsize}{!}{\includegraphics{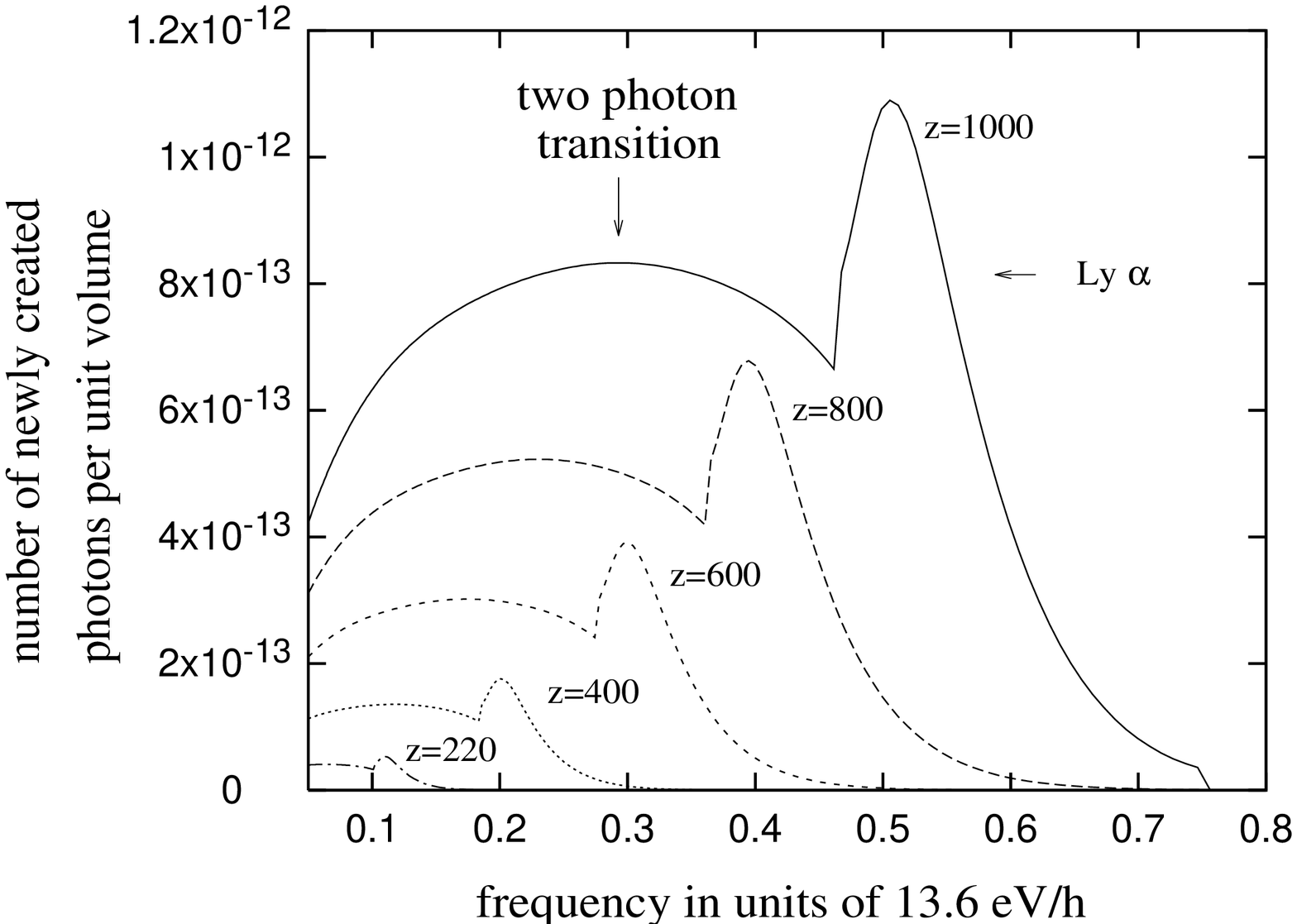}}
\caption{Spectra of recombination--photons 
at different redshifts
for  $\Omega = 1$,$\, \Omega_b = 0.1$,$\, h = 1$.}
\label{Fig4}
\end{figure}

\begin{figure}[htbp]
\resizebox{\hsize}{!}{\includegraphics{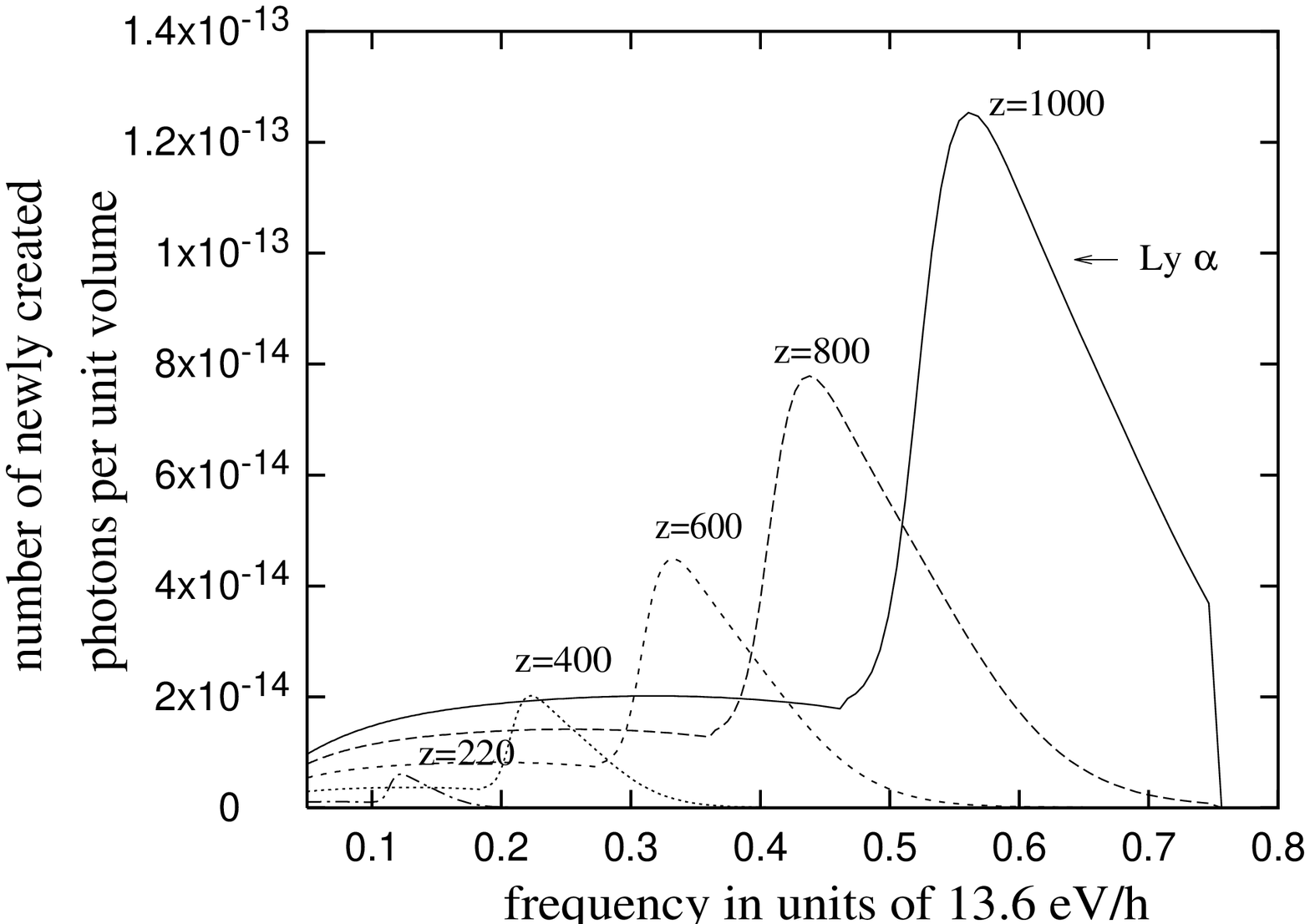}}
\caption{Spectra of recombination--photons 
at different redshifts
for  $\Omega = 1$,$\, \Omega_b = 0.01$,$\, h = 1$.}
\label{Fig5}
\end{figure}

\begin{figure*}[htbp]
\resizebox{\hsize}{!}{\includegraphics{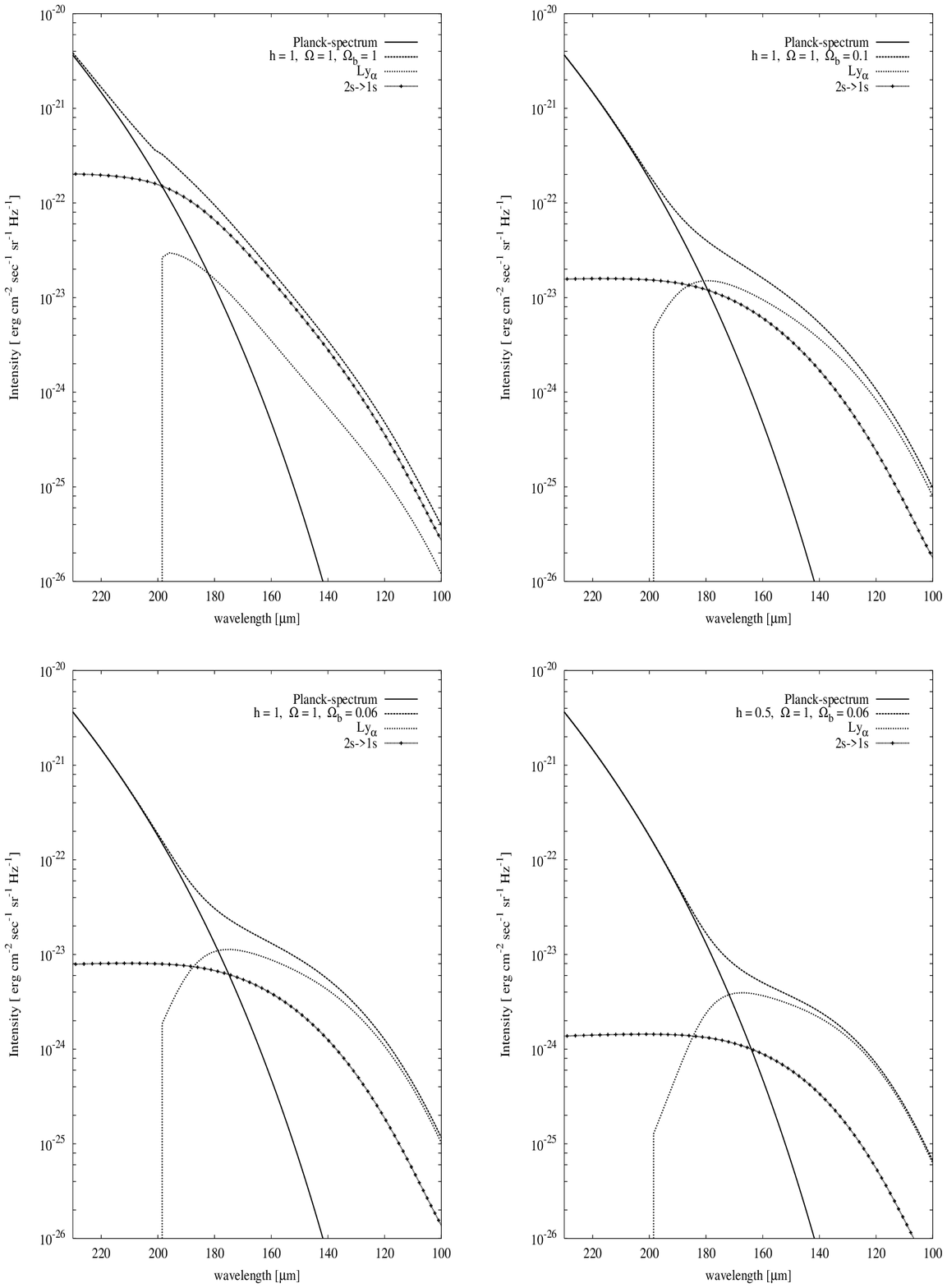}}
\vspace{-3.2cm}
\caption{Contribution of the Ly$_{\alpha}$ Photons and the
 $2s \rightarrow 1s$ two--photon transitions to the distortion of the CMB for 
different cosmological models. }
\label{Fig6}
\end{figure*}

\subsection{The spectrum of photons issuing from the recombination}

From the solution of equation (\ref{receq}) we know the 
time dependence of the free electron density ${n_e(t)}$. Using the equations
(\ref{rph2}) and (\ref{rsr})  the net two--photon transition rate and the 
${ Ly}_{\alpha}$ redshift rate can be calculated. The spectrum of 
the photons emerging from the $2s \rightarrow 1s \,$ transition
can be taken from  Spitzer \&  Greenstein (1951). The appropriately normalized
two--photon spectrum in terms of the  variable $x$ is:
\begin{equation}  
\label{ph2s}
        \Phi (x) = 0.7081 e^x 
        \psi \biggl( {4 \over 3} e^x\biggr). 
\end{equation}
The $ \int^{x_{12}}_{-\infty}\Phi(x) d x = 2$. 

For our purposes the frequency distribution of the ${\rm Ly}_{\alpha} \,$ 
photons  can be approximated by a delta function on the $\nu$ axis, so 
\begin{equation}
   Q(x,\tau) = R_{2ph} \Phi(x)+ R_{rs} e^{x - x_{12}} \delta (x - x_{12})\, .
\end{equation}

\begin{figure*}[htbp]
\resizebox{16.5cm}{16.5cm}{\includegraphics{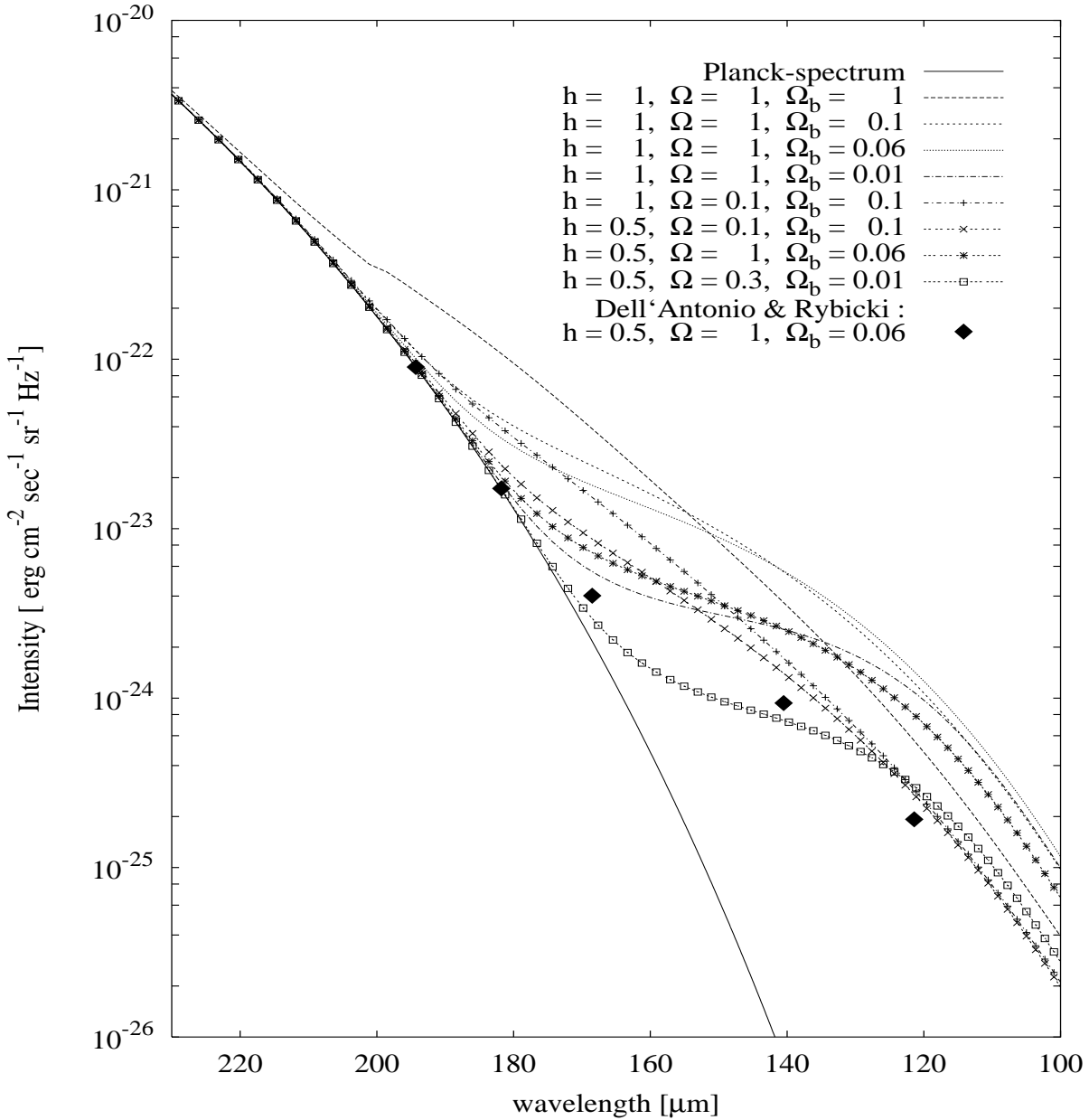}}
\caption{CMB Spectra for different cosmological models.}
\label{Fig7}
\end{figure*}

\subsection{The dependence of the spectrum on cosmological parameters}

With $n_{1s}(t)$, the ratio of ${\rm Ly}_{\alpha}$ redshift rate to the 
two--photon decay rate  can be calculated from equations (\ref{rsr}) and
(\ref{rph2}). This ratio depends on the cosmology. For the $2 \rightarrow 1 \,$
transition neglecting the $-1$ in the denominator of equation (\ref{rsr}) and 
using the numerical values, one arrives at
\begin{equation}
\label{rat}
        R = {R_{rs} \over R_{2phot}} = {8 \pi  \nu_{12}^3  c^{-3} H \over
         A_{2s,1s} n_{1s}}
         = {1 \over 1 - x} \;  {632.92 \over (1 + z)^{3/2}}  
        {\Omega^{1/2} \over \Omega_b h}.              
\end{equation}

The redshift of the new--born photons is given by equation (\ref{lapleq}) in 
terms of the variables $x$ and $\tau$. The spectrum at different redshifts is 
shown in Figure 4 ($ \Omega = 0.1$, $\Omega_b =0.1$, $h=1$) and 
Figure 5 ($ \Omega = 1$, $\Omega_b =0.01$, $h=1$). When the baryon 
density is low, the 2s $\rightarrow $ 1s transition plays a minor role.

These results  disagree with Dell'Antonio \& Rybicki (1993), who state that the
energy distribution of the photons emitted  by the $2s \rightarrow 1s$ 
transition rather strongly peaked at $\nu =n_{{\rm Ly}_{\alpha}}/2$  and
the $2s\rightarrow 1s$ transitions give no more than 1\% difference
in the free electron densities and the line strength. Our results 
(Fig. 4 and Fig. 5) show that the spectrum of $2s\rightarrow 1s$ photons is
broad (see also Spitzer \& Greenstein 1951). Though  for certain  
cosmologies the contribution of these photons  is small, but as it is 
demonstrated in Figure 6, for other set of parameters this
 contribution is quite considerable.  
In the models with high $\Omega$ and low $\Omega_b$ the
 two--photon transitions are negligible, but for $h^2 \Omega_b \ge 0.1$ they
 are more important, and for the flat, high $\Omega_b$ model they are
 dominant. In spite of the
 above mentioned disagreement our recombination history curve
(Fig. 2) agrees well with that of  Dell'Antonio \& Rybicki (1993), because by
computing this function they took into account also the 2s $\rightarrow$ 1s 
transitions.

Figure 7 shows  a part of the CMB radiation
spectrum  for different cosmological parameters, where the distortion due to
recombinations is the most important. 
The hydrogen recombination begins when in the background radiation there are 
less energetic photons than hydrogen atoms. Consequently, the spectrum of the 
photons issuing from the recombination has a maximum near the Planck-curve.
The other maximum in this spectrum corresponds to the two--photon 
transitions, is longwards from the first maximum at $h \nu_{{\rm max},2} =
(B_2 - B_1)/2 = 5.1$ eV, and is under the Planck--curve. The $2s 
\rightarrow 1s$ photons
influences the distortion in two ways. First, the short wavelength part of
 their spectrum is above the Planck-curve and give a 
direct contribution to the distortion (Figure 6). Second, the number of 
recombinations are given by the number of hydrogen atoms. If a considerable 
part of recombinations happens by two--photon decays, there are less 
redshifted Ly$_{\alpha}$ photons, the amplitude and shape of their
 spectrum changes.

For given 
$\Omega_b\,$ and h there are more photons above the Planck--curve when both
$\Omega \,$  and R are large. When $\Omega\,$ and h are  given,
the spectrum with the larger $\Omega_b \,$ lies above all the others. For a 
fixed value of  $\Omega$ and $\Omega_b \,$ the spectrum with larger $h$
 is larger. 

In Figure 7 we compare our spectrum with that of Dell'Antonio \& Rybicki 
(1993). In the 120 $\mu$m $< \lambda <$ 170 $\mu$m range there are about a 
factor of 2--3 more  photons as seen by comparing our curve with the diamonds.
That may be caused by the different technique in computing the recombination
process. It seems us somewhat arbitrary to handel the 10 lowest level in a
different way as the others. Burgess (1958) used this method to compute the 
recombination spectrum  in nebulae. In the case of nebulae this method
gives good results, because there is no thermal 
equilibrium above the n = 2 states, the electrons recombine on the 
low lying states and cascade down.

\section{Summary and Discussion}

Background photons stimulate recombination processes and that give an
important contribution to the effective recombination coefficient. 
The resulting 
recombination coefficient is about a factor of four larger than calculated
without the induced recombinations (Figure 1) and  depends, besides the 
temperature, on the free electron (proton) density as well. 
In contrast to the recombining
Universe in a gaseous nebulae with the same temperature and free electron 
density the recombination coefficient is given by  (\ref{alpeq}), because 
there are no photons that should stimulate recombination on excited states. 
The photons, except the Ly$_{\alpha}$ photons, emerging from recombination in the nebulae leave the nebula without interaction.

For any given principal quantum number the $\alpha_{nl}\,$ recombination 
coefficients  depend considerably on the angular momentum quantum number 
$l$. There is a peak around $l = 4$ and  $\alpha_{nl}\,$ vanishes when $l$ is 
high. Transitions 
between states with high $l$'s are slow, which could influence the 
relative occupation of s and p states. Since this last effect is probably 
small (see Hu et al. 1995), we have completely neglected it here.

In the main part of the recombination process the  free--bound 
transitions are more rapid than the depopulation of the  $n = 2$ state and
the thermodynamic equilibrium  is maintained by reionization. Consequently, 
the course of the recombination process does not depend on the details of the
$\alpha ({T,n_e}) \,$ function. Because of that, authors, using quite different
effective recombination coefficients, come up with similiar time dependence 
for recombination. When, however, the free electron density
is already low, the depopulation of the $n = 2$ states is more rapid than the 
recombinations of the $n \ge 2$ states and
the speed of the recombination process is determined by the
free--bound transitions of the excited states and the induced recombinations 
are important.  As a consequence, our ionization curve (Figure 2) is for 
small ionization grades steeper than that of Dell'Antonio \& Rybicki (1993), 
who did not take  this effect into account.

The last scattering of the CMB photons happens, with high probability, in the
first phase of the recombination, when the number of free--bound transitions 
to the excited states is large. The induced recombinations and
the numerical value of the effective recombination coefficient do not 
influence the position of the last scattering layer. Our results agree with
those of White, Scott \& Silk (1994).

On the other hand the residual ionization is a quantity determined by the 
details of the recombinations of the excited states at low n$_e$. 
At $z < 500 \,$, whether induced recombination is taken into account or not,
the fractional ionization is proportional to 
$\sqrt{\Omega} \Omega_b^{-1} h^{-1}$. 
However, because of the induced recombinations, the residual ionization
is reduced by a factor of about 2.6. The exact value of this factor depends on
the cosmological parameters.
A consequence of the low residual ionization and the higher number of photons 
in the tail of the CMB spectrum is that the final abundance of molecular 
hydrogen will be much smaller than that estimated by Lepp \& Shull (1983).

The hydrogen recombination gives  contribution to the CMB spectrum
in the $\lambda <$ 190 $\mu$m range. The amplitude and form of this distortion 
depends on the number of photons issuing from the recombination, and their
distribution between the two decay way, the $2s \rightarrow 1s$ two--photon 
mode and the redshifted  Ly$_{\alpha}$ mode. The induced recombinations
only in the last phase influence the recombination process, when the 
ionization grade already very low. This gives a very small contribution to the 
distortion of the spectrum.

The contribution of helium recombination to the photon spectrum was 
discussed by Lyubarsky \&
Sunyaev (1983) and by Fahr \& Loch (1991). According to their results
the recombination of 
He sets in at a redshift of about four times larger than that for hydrogen 
recombination. But the photon energy is also four times larger and
the photons emerging from helium recombination could also distort the CMB
spectrum in the same frequency range as the H--recombination. 
However, as it was pointed out by Peebles (1995), at those redshifts where  
He II recombines there is already a trace of recombined hydrogen, which 
can absorb some of the photons created during helium recombination.
Thus direct recombination into 
the ground state is possible for helium and the Saha formula is a good
approximation to describe the helium recombination history. 
The optical depth due to ionization between z$_1$ and z$_2$ is
\begin{equation}  
\label{Heph}
        \tau = \int_{z_1}^{z_2} \sigma_{ion} n_{1s} {d r \over d z} d z\,.
\end{equation}  
The redshift of the helium recombination is z$_1$, the  cross section for
hydrogen ionization is $\sigma_{ion}\, [ \approx (B_1/h \nu)^{8/3}, {\rm with}
\, \nu(z) = \nu(z_1) {z \over z_1}$], and  the number
density of hydrogen atoms, as computed from the Saha formula, is n$_{1s}(z)\,$.
The distance to the radiation source at the moment of emission, r(z$_1$), is
calculated from the Friedmann equation. The integral reaches  1 at about z$_2 \approx 1900$. Consequently,
in contrast to the results mentioned above, helium recombination does not leave any trace on the CMB spectrum.

As is discussed in Hu et al. (1995), the difference between the temperature of
the  photons
and the kinetic temperature of the electrons has an effect below  z
$\approx$ 200. We follow the recombination process only down to z = 220, 
so this  effect is neglected.

The possible $ns \rightarrow 1s\,$ and the  $nd \rightarrow 1s\,$ two--photon
transitions could also have some  importance. The number of bound electrons
in the $n \ge 3$ states is of the same order of magnitude that for n$_{2s}$. 
The two--photon decay probabilities for these states are smaller than for $n_{2s}$.
Moreover some of these photons will have more energy than  B$_2$ -- B$_1$ =
10.2 eV. However, the contribution of these transitions is not entirely 
negligible, they can change the balance between the two routes, redshift and 
two--photon decay, for the elimination of energetic photons and change the 
spectrum of the recombination photons.
This correction will be discussed in detail elsewhere.

If there
were some energy input prior to the hydrogen recombination and a
"y-distortion" of the spectrum (Zel'dovich \& Sunyaev 1969) it would be 
easier to observe the consequences of H--recombination (Lyubarsky \& 
Sunyaev 1983). However, the observations permit a very low upper limit  
for the y-distortion (Mather et al.\ 1994, Fixsen et al.\ 1996).

The distortion seems to occur
at wavelengths  where other sources give considerable contributions.
By doing the calculations we  had hoped that
the zodiacal forground emission, the dust and molecular emission from the 
interstellar medium would be reliably modeled, subtracted and the
calculated spectral distortion would be measurable.
Recently that substraction has been done (Schlegel, Finkbeiner \& Davis 1997) 
and it turned out that in the considered wavelength range 
the measured upper limit of the background is at least two order of magnitude
 higher as the spectrum calculated here. As it was pointed out by Guideroni
 et al. (1997), it could originate from early starlight scattered by dust.
\begin{acknowledgements}
We thank Chris Tout for carefully reading the manuscript. P\'eter Bosch\'an
wishes to thank David Hummer for valuable discussion and advising. 
\end{acknowledgements}


\begin{thebibliography}{}
\bibitem[]{} Boardman, W.\ J., 1964, ApJS  9, 185
\bibitem[]{} Burgess, A., 1958, MNRS, 18, 477
\bibitem[]{} Dell'Antonio, J.\ P. Rybicki, G. B., 1993, in
Observational Cosmology, ASF conference Series Vol. 51 
\bibitem[]{} Fixsen, D.\ J., Cheng, E.\ S., Gales, J.\ M., Mather, J.\ C.,
        Shafer, R.\ A. \& Wright, E.\ L., 1996, ApJ 473, 576
\bibitem[]{} Guideroni, B., Bouchet, F. R., Puget, J.-L., Lagache, G. 
\& Hivon, E., 1997, Nature 390, 257
\bibitem[]{} Hu, W., Scott, D., Sugiyama, N. \& White, M., 1995, Phys. Rev. D 
52, 5498
\bibitem[]{} Hummer, D.\ G., 1994, MNRS 268, 109
\bibitem[]{} Hummer, D.\ G. \& Mihalas, D., 1988, ApJ 331, 794
\bibitem[]{} Fahr, H.\ J. \& Loch, R., 1991, A\&A 246, 1
\bibitem[]{} Krolik, J.\ H., 1989, ApJ 338, 594
\bibitem[]{} Krolik, J.\ H., 1990, ApJ 353, 21
\bibitem[]{} Jones, B.\ J.\ T. \& Wyse, R.\ F.\ G., 1985, A\&A 149, 144
\bibitem[]{} Lepp, S. \& Shull, J. M., 1984, ApJ 280, 465
\bibitem[]{} Lyubarski, Y.\ E. \& Sunyaev, R.\ A., 1983, Ap\&SS 123, 171
\bibitem[]{} Matsuda,T., Sato, H. \& Takeda, H., 1971, 
Progr. Theor. Phys. 46, 216
\bibitem[]{} Mihalas, D. 1984, Foundation of Radiation Hydrodynamics, 
        Oxford University Press, N.Y. 
\bibitem[]{} Milne E. A., 1924, Phil. Mag. 47, 209
\bibitem[]{} Novikov, I.\ D. \& Zel'dovich, Ya.\ B., 1967, 
        The Structure and Evolution of the Universe, 
        The University of Chicago Press, Chicago and London

\bibitem[]{} Peebles, P.\ J.\ E., 1968, ApJ 153, 1

\bibitem[]{} Peebles, P.\ J.\ E., 1993, Principles of Physical Cosmology, 
        Princeton University Press
\bibitem[]{} Peebles, P.\ J.\ E., 1995 Quoted as privat communication in
Hu et al. 1995.

\bibitem[]{} P\'equignot, D., Petitjean, P. \& Boisson, C.,1991, A\&A 251, 680 
\bibitem[]{} Puget, J.-L., Abergel, A., Bernard, J.-P., Bulanger F., 
Burton, W. B., D\'esert, F.,-X.  \& Hartmann, D., 1996, A\&A 308, L5 
\bibitem[]{} Rybicki, G.\ B. \& Dell'Antonio, I.,\ P., 1994, ApJ 427, 603
\bibitem[]{} Rybicki, G.\ B. \&  Lightman, A.\ P., 1979, Radiative 
        Processes in Astrophysics, John Wiley New York

\bibitem[]{} Sasaki, S. \& Takahara, F., 1993, PASJ 45, 655
\bibitem[]{} Schlegel, D.J., Finkbeiner, D.P. \& Davis, M., 1997, BAAS 191,
87.04 (astro-ph/9710327)
\bibitem[]{} Spitzer, L.\ A. \& Greenstein, J.\ L., 1951, ApJ 114, 407
\bibitem[]{} Storey, P.\ J. \& Hummer, D.\ G., 1991, Computer Physics 
        Communications  66, 129
\bibitem[]{} White, M., Scott, D. \& Silk, J., 1994, Annu. Rev. Astron. 
Astrophys. 32, 319

\bibitem[]{} Zel'dovich, Ya.\ B., Kurt, V.\ G. \& Sunyaev, R.\ B., 1968,
        Zh. Eksp. Teor. Fiz. 55, 278,\\ 
        englisch translation: 1969, JETP 28, 146

\bibitem[]{} Zel'dovich, Ya.\ B. \& Sunyaev, R.\ B., 1969, Ap\&SS 4, 301.
\end{thebibliography}
\end{document}